\documentclass[sigconf,screen]{acmart}
\AtBeginDocument{%
  \providecommand\BibTeX{{%
    \normalfont B\kern-0.5em{\scshape i\kern-0.25em b}\kern-0.8em\TeX}}}
    \makeatletter
\def\@copyrightpermission{\relax}
\makeatother
\acmDOI{}
\acmISBN{}
\usepackage{graphics}
\usepackage{epsfig} 
\usepackage{caption}
\usepackage{subcaption}

\begin{document}

\title{A Novel U-Net Architecture for Denoising of Real-world Noise Corrupted Phonocardiogram Signal}

\author{Ayan Mukherjee}

\affiliation{
  \institution{TCS Research \\ Tata Consultancy Services}
  \city{Kolkata}
  \state{West Bengal}
  \country{India}}
  \email{ayan.m4@tcs.com}
  
\author{Rohan Banerjee}
\affiliation{%
  \institution{TCS Research \\ Tata Consultancy Services}
  \city{Kolkata}
  \state{West Bengal}
  \country{India}}
\email{rohan.banerjee@tcs.com}

\author{Avik Ghose}
\affiliation{%
  \institution{TCS Research \\ Tata Consultancy Services}
  \city{Kolkata}
  \state{West Bengal}
  \country{India}}
\email{avik.ghose@tcs.com}







\begin{abstract}
 The bio-acoustic information contained within heart sound signals are utilized by physicians world-wide for auscultation purpose. However, the heart sounds are inherently susceptible to noise contamination. Various sources of noises like lung sound, coughing, sneezing, and other background noises are involved in such contamination. Such corruption of the heart sound signal often leads to inconclusive or false diagnosis. To address this issue, we have proposed a novel U-Net based deep neural network architecture for denoising of phonocardiogram (PCG) signal in this paper. For the design, development and validation of the proposed architecture, a novel approach of synthesizing real-world noise corrupted PCG signals have been proposed. For the purpose, an open-access real-world noise sample dataset and an open-access PCG dataset has been utilized. The performance of the proposed denoising methodology has been evaluated on the synthesized noisy PCG dataset. The performance of the proposed algorithm has been compared with existing state-of-the-art (SoA) denoising algorithms qualitatively and quantitatively. The proposed denoising technique has shown improvement in performance as comparison to the SoAs.
\end{abstract}

\ccsdesc[500]{Computing methodologies}
\ccsdesc[500]{Computing methodologies~Neural networks}
\ccsdesc[500]{Applied computing}
\ccsdesc[500]{Applied computing~Life and medical sciences}
\ccsdesc[500]{Applied computing~Health informatics}

\keywords{Heart sound, Deep learning, Real-world noise, Denoising architecture, Phonocardiogram, U-Net}



\maketitle
\section{introduction}

\begin{figure}
	\begin{subfigure}{.48\textwidth}
		\centering
		\includegraphics[width=0.96\textwidth]{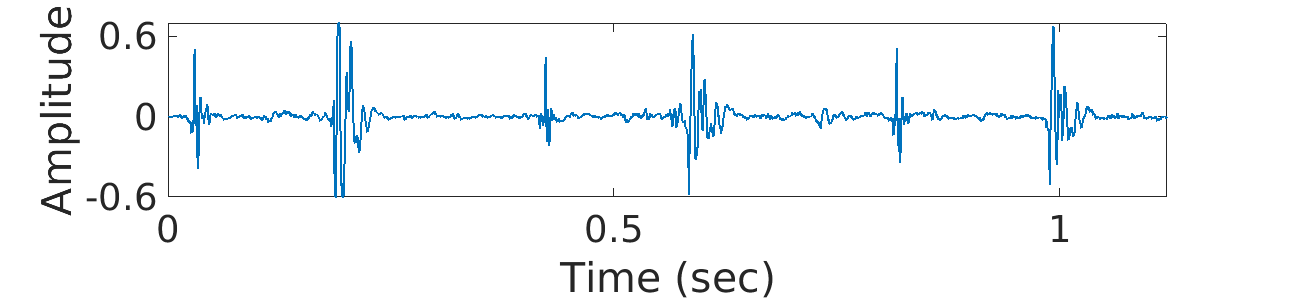}
		\caption{Plot of clean phonocardiogram signal}
		\label{fig:clean_normal.png}
	\end{subfigure}
	\begin{subfigure}{.48\textwidth}
		\centering
		\includegraphics[width=0.96\textwidth]{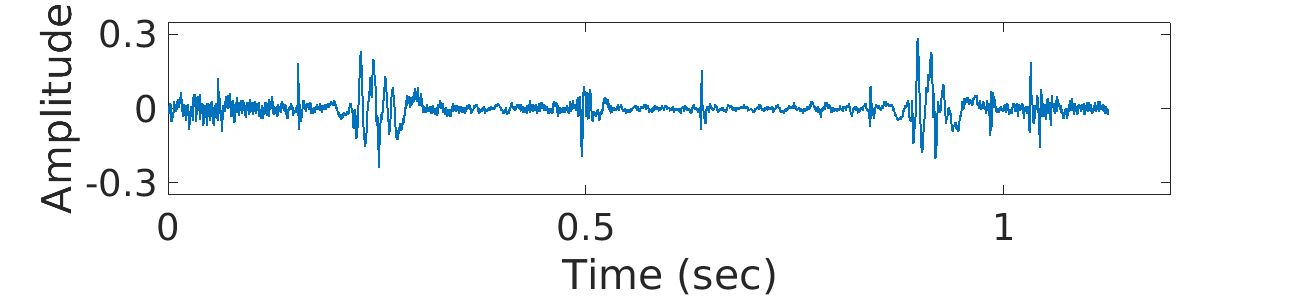}
		\caption{Plot of noisy phonocardiogram signal}
		\label{fig:clean_murmur.png}
	\end{subfigure}
	\caption{Plot of noisy and clean phonocardiogram signal}
	\label{fig:Typical real-world clean and noisy phonocardiogram signals}
\end{figure}

The pumping action of the heart circulates blood throughout the body. During the circulation the opening and closing of the heart valves gives rise to the heart sounds. The four fundamental heart sounds are: 1) first heart sound (S1), 2) second heart sound (S2), 3) systolic interval and 4) diastolic interval. The heart sound is heard by physician/clinician using stethoscope for auscultation purposes. However, owing to the characteristic low amplitude of the heart sound signal, it is naturally susceptible to ambient noises \cite{leal2018noise}. 
Sample recordings of a noisy and a clean PCG signals are plotted in Fig. \ref{fig:clean_normal.png} and Fig.~\ref{fig:clean_murmur.png} respectively. 
It can be surmised from the figures that in the case of the noisy heart sound recording, the significant features of the signal become obfuscated due to the noise corruption. Under such a scenario, reliable auscultation become difficult even for the experts at places like out patients departments and non-clinical environments.

Thanks to the recent advances in artificial intelligence and machine learning techniques, an automatic diagnosis of different pathological conditions is possible from a PCG.  However, noisy signals severely impact the performance of such algorithms as well. Hence, it can be concluded that denoising is an essential and practical pre-processing step required to ensure reliable auscultation/decision-making for human experts/machine driven algorithms. 

The research complexity of denoising PCG corrupted with real-world noise arises due to 1) the wide gamut of naturally occurring noise sources that can corrupt the heart sound, and 2) the significant spectral overlap that exists between the heart sound spectrum and the noise spectrum. 

\begin{figure}[]
	\centering
	\includegraphics[width=0.48\textwidth]{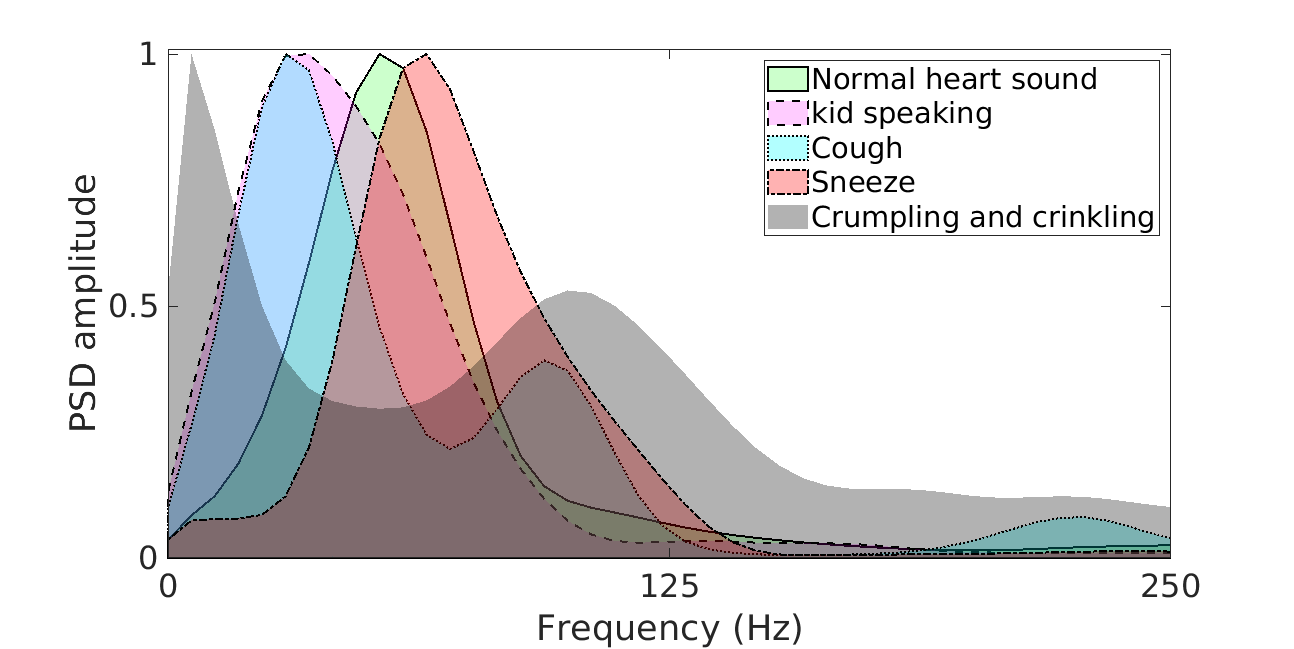}
	\caption{Spectrum of PCG signal and typical real-world noises}
	\label{Frequency_overlap_HS_noise}
\end{figure}
In Fig.~\ref{Frequency_overlap_HS_noise} , the spectral plot of typical heart sound cycles, and other real-world noise samples like \textit{child speech}, \textit{sneeze}, \textit{cough}, \textit{crumpling and crinkling} are plotted. The significance of overlap between the spectrums as can be observed from the figure. 




Two types of approaches for dealing with noisy heart sound signals exist in literature. In the first approach, the noisy part of the signal is identified and discarded. Such noisy segment identification is done based on signal quality metrics \cite{springer2014signal}. The other approaches modifies the noisy signal through some form of filtering like bandpass filtering \cite{de2007automated}, spectral subtraction \cite{ tosanguan2008modified, paul2006noise}, etc. to retrieve the clean signal. However, due to the spectral overlap between the signal and noise spectrum such methods are largely ineffective \cite{asmare2021can}. Wavelet transform is another widely used technique used for denoising of heart sound \cite{ agrawal2013wavelet, messer2001optimal}. However, the performances of such algorithms are highly sensitive to the setting of the thresholds \cite{asmare2021can}. 

While all the existing state-of-the-art (SoA) heart sound denoising algorithms have reported effective denoising of  noisy heart sound signals, most of those have considered only additive white gaussian (AWG) signal as the noise component \cite{asmare2021can}. This makes the reliability of such PCG denoising algorithms uncertain under real-world noise corruptions. 

In order to address this research issue, the present research work proposes a deep learning architecture for denoising of PCG signals corrupted with real-world noise recordings.

The major contributions of the present research work are:
\begin{enumerate}
    \item Development of a noisy PCG signal dataset based on real-world noise samples (child speak, cough, sneeze, crumpling and crinkling, hiss).
    \item Development of a U-net based deep learning denoising architecture for reliable denoising of real-world noise corrupted heart sound signals.
    \item Thorough evaluation of the proposed denoising architecture with simulated real-world noise corrupted PCG signal dataset as well as performance comparison with the existing SoA denosing algorithms.
 
\end{enumerate}

The rest of the paper is organized as follows:

Section II presents the process for the synthesis of the real-world noise corrupted PCG signal dataset. Section III provides a detailed description of the proposed U-Net based denoising architecture. The evaluation of the proposed architecture and comparative analysis with existing SoA denosing techniques are presented in Section III followed by conclusion in Section IV.
\section{dataset description} 
\subsection{Heart sound dataset}
For the present research work, a subset of the publicly available \textit{PASCAL} heart sound dataset (\textit{Btraining\_normal} subset) \cite{pascal-chsc-2011} has been utilized. The choice of the dataset was motivated by the availability of clean heart sound signals, which is the fundamental requirement of the present research endevour. The dataset consists of $200$ clean heart sound signals of varying lengths (between $1$ second and $30$ seconds). The signals are sampled at $4$ KHz and saved as audio files in \textit{wav} format.
\subsection{Real world noise sample dataset}
In order to generate the noisy PCG recordings portions of real-world noise recordings have been used. The publicly available \textit{ARCA23K} dataset \cite{Iqbal2021} provides labeled real-world sound events. It consists of $23727$ audio clips of varying lengths and along with annotations ($70$ class labels). All the \textit{ARCA23K} recordings are sampled at $44.1$ KHz and saved as audio files in \textit{wav} format.
\subsection{Noisy heart sound dataset synthesis}
\begin{figure}
	\begin{subfigure}{.48\textwidth}
		\centering
		\includegraphics[width=0.96\textwidth]{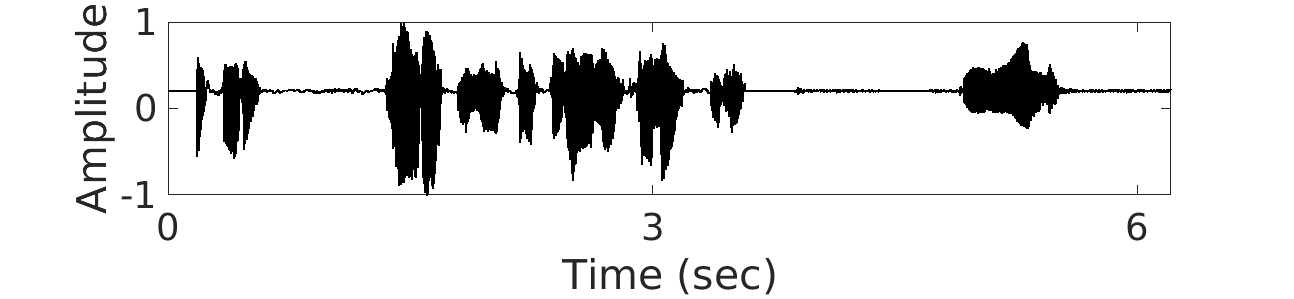}
		\caption{\textit{Child speech} noise vector}
		\label{fig:Child_speech}
	\end{subfigure}
	\begin{subfigure}{.48\textwidth}
		\centering
		\includegraphics[width=0.96\textwidth]{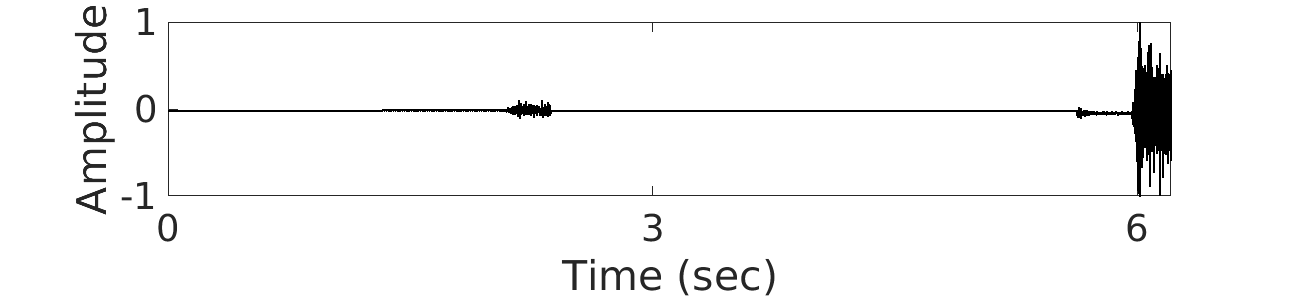}
		\caption{\textit{Cough} noise vector}
		\label{fig:cough}
	\end{subfigure}
	\begin{subfigure}{.48\textwidth}
		\centering
		\includegraphics[width=0.96\textwidth]{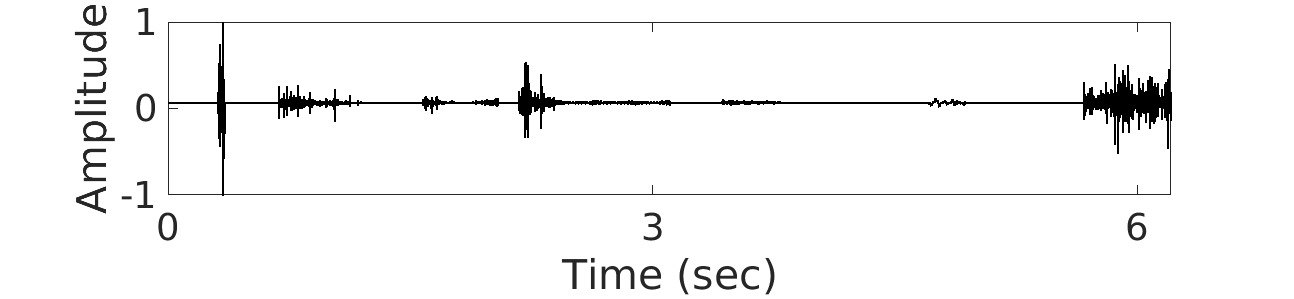}
		\caption{\textit{Crinkling and crumpling} noise vector}
		\label{fig:Crinkling}
	\end{subfigure}
	\begin{subfigure}{.48\textwidth}
		\centering
		\includegraphics[width=0.96\textwidth]{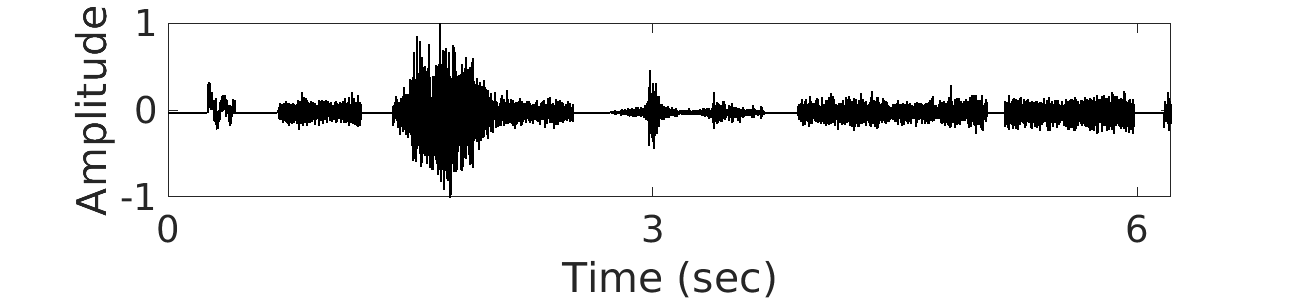}
		\caption{\textit{Hiss} noise vector}
		\label{fig:Hiss}
	\end{subfigure}
	\begin{subfigure}{.48\textwidth}
		\centering
		\includegraphics[width=0.96\textwidth]{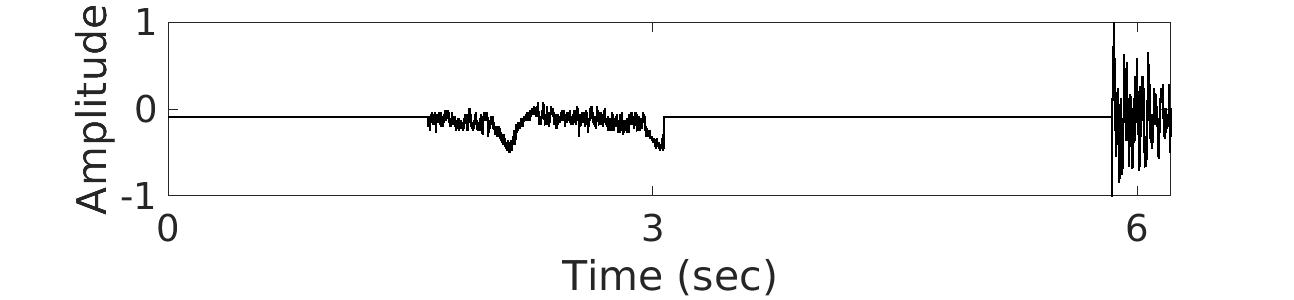}
		\caption{\textit{Sneeze} noise vector}
		\label{fig:Sneeze}
	\end{subfigure}
	\caption{Samples of normalized noise vectors for different labels}
	\label{fig:Noise_samples}
\end{figure}

Among the $70$ labels of the \textit{ARCA23K} dataset, only a subset of it has relevance (can be considered as noise) for heart sound corruption in a real-world setting. Hence, for the present work, we considered only those audio files that had one of the following labels as annotation: (1) \textit{Crumpling and crinkling}, (2) \textit{Child speech and kid speaking}, (3) \textit{hiss}, (4) \textit{sneeze}, and (5) \textit{cough}. 
All such relevant noise recordings were resampled to $4$ KHz to match the sampling rate of the \textit{PASCAL} heart sound recording. Now, the steps followed for mixing the down-sampled noise recordings with the clean PCG signals are enumerated as follows: 
\begin{enumerate}
	\item Random number of segments of varying lengths are generated from noise audio samples.
	\item For each category of noise, the segments are randomly placed on a zero vector of length equal to the heart sound recording under consideration. This results in the generation of category-wise noise vectors.
	\item The noise vectors are normalized between $[-1,1]$.
	\item Now, the noise vectors are added sample-wise to the heart sound vector.
	\item The resultant vector is again normalized between $[-1,1]$ to generate the noisy heart sound vector.
\end{enumerate}
\begin{table}[]
	\caption{Noise statistics}
	\label{Noise_statistics}\centering
\begin{tabular}{p{3.7cm}p{1cm}p{0.75cm}p{0.75cm}p{0.75cm}}
Noise category & \# of segments & Mean (Sec) & Std (Sec) & Mode (Sec) \\ \hline
\textit{Child speech and kid speaking} & 5892           & 0.76       & 0.76                                                & 0.29       \\ \hline
\textit{Hiss}                     & 21719          & 0.64       & 0.68                                                & 0.37       \\ \hline
\textit{Crumpling and crinkling}        & 21569          & 0.62       & 0.66                                                & 0.11       \\ \hline
\textit{Cough}                     & 5910           & 0.84       & 0.73                                                & 0.45       \\ \hline
\textit{Sneeze}                      & 5892           & 0.84       & 0.73                                                & 0.25  \\ \hline 
\end{tabular}
\end{table}
Samples of the noise vectors generated following the above steps are shown in Fig.~\ref{fig:Noise_samples}.
\section{methodology}
\subsection{Pre-processing}
\begin{figure}
	\begin{subfigure}{.45\textwidth}
		\centering
		\includegraphics[width=0.8\textwidth]{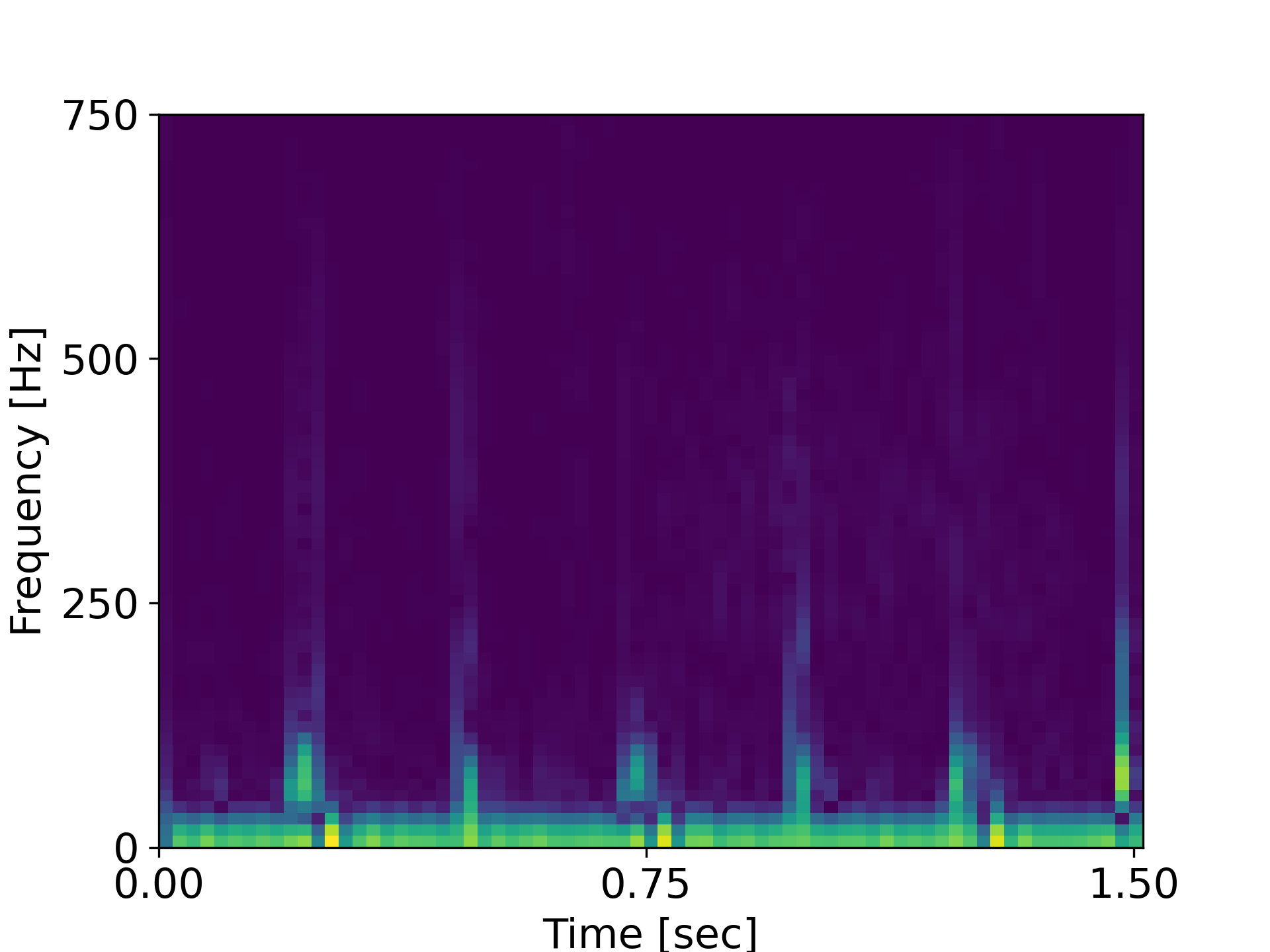}
		\caption{STFT plot for a dataframe of clean a PCG signal}
		\label{fig:clean_stft.png}
	\end{subfigure}
	\begin{subfigure}{.45\textwidth}
		\centering
		\includegraphics[width=0.8\textwidth]{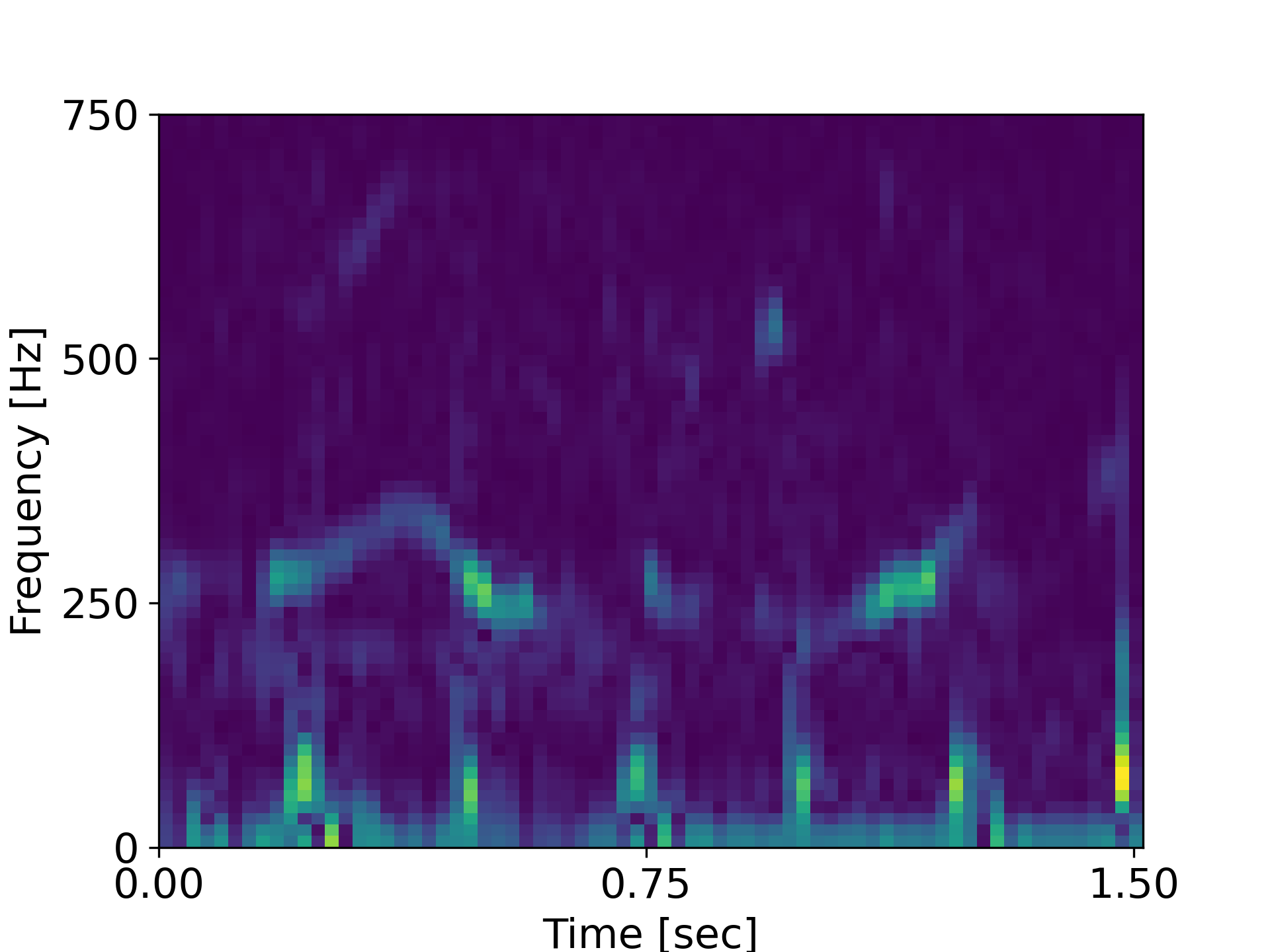}
		\caption{STFT plot for a dataframe of a noisy PCG signal}
		\label{fig:Noisy_stft.png}
	\end{subfigure}
	\caption{STFT plots corresponding to dataframes ($1.5$ seconds) of clean and noisy PCG signals}
	\label{fig:typical clean and noisy STFT}
\end{figure}

Taking cognizance of the limited spectral bandwidth of a typical heart sound cycles (Fig. \ref{Frequency_overlap_HS_noise}) each noisy PCG time series is down-sampled from $4$ KHz to $1500$ Hz. Next, from the down-sampled signal, non-overlapping data-frames are extracted (window length = 1.5 seconds, rectangular window). Now each such 1-D dataframe is transformed to the time-frequency domain using short time Fourier transform (STFT). The STFT parameters used for the present application (samples per segment = $64$, FFT points = $128$, window = $Hanning$, scaling  = $spectrum$, segment overlap= $50\%$). These are standard STFT parameters reported in research literature \cite{grais2017two}.  This transformation generates the spectrum matrix ($Z_{xx}$) with dimension $65\times72$. The typical STFT frames generated for clean and noisy dataframes are shown in Fig.~\ref{fig:clean_stft.png} and Fig.~\ref{fig:Noisy_stft.png} respectively. Clear reflection of the noise corruption can be observed in Fig.~\ref{fig:Noisy_stft.png}. $Z_{xx}$ is further split into its real and imaginary matrix components. The component matrices are resized to $64\times64$  and are concatenated along the third axis. The resulting $64\times64\times2$ matrix is fed into the deep neural network architecture discussed in the following section. 
\subsection{Proposed U-net based denoising approach }
U-Net is a widely used fully convolutional deep neural network architecture. It has been widely applied in image segmentation \cite{ronneberger2015u}, image denoising \cite{bao2020real} and restoration \cite{aghabiglou2021projection}. However, to the best of the knowledge of the authors, the present application of the U-Net architecture for the denoising of 1D physiological time series data is a novel application of U-Net. The U-Net has a similar architecture to that of the denoising autoencoder-decoder network. However, in U-Net, the presence of skip connections transfer the fine-grained information from the  analysis path to the synthesis path. Such information allows the network to recostruct signals with accurate finer morphological details. This property of the U-Net is the motivation behind its choice as the learning model for the present application. For the purpose, the architecture proposed in  \cite{ronneberger2015u} has been utilized with minor modifications. The architecture can be referred to in \cite{ronneberger2015u}.
The flow of the U-net architecture in the context of the present application is summarized as follows:

 The $64\times64\times2$ input (as discussed in the preceeding subsection) is filtered by 2D convolution filters (number of filters are increased from $8$ to $128$ in steps) and downsampled using \textit{max-pooling}($2\times2$) in steps to $4\times4\times128$. This latent space is again filtered and upsampled for signal synthesis such that the output of the final layer dimension-wise matches the input. For the optimization of the tunable network parameters Nadam optimizer has been employed ($\beta_{1}$ = $0.9$,$\beta_{2}$ = $0.99$,$\epsilon$ = $1e-07$ , learning rate = $0.0005$). The optimized function is the mean-squared error (\textit{mse}). Further \textit{early-stopping criteria} has been imposed upon the training to mitigate the curse of over-fitting. The batchsize is set to 128 and maximum epoch is set to 100. Further, batch normalization has been employed to limit the effect of over-fitting. Rectified linear unit (ReLU) function has been used as the non-linear activation function across all the layers of the network.   
\subsection{Post-processing}
In the post-processing stage, the ouput spectrum frames are  resized to ($65\times72$) and then transformed back to the 1-D data frames using inverse STFT. Finally the frames are concatenated to obtain the denoised heart sound time series.
\section{Results and discussion}
The proposed architechture has been simulated in the Python environment using Pycharm IDE. The TensorFlow version used for the U-Net architecture realization is 2.7.4. From the $200$ clean \textit{PASCAL} heart sound recordings, following the process already discussed (\textit{section II}, \textit{subsection C}), 4000 noisy recordings has been generated. The noisy recordings is further split into training, validation and test partitions in the ratio of 64:16:20. The split has been done at the subject level in order to ensure that data of none of the subjects is present in more that one partion. The real-world noisy heart sound data simulation and the deep learning training and testing is done in a computing device with 16 GB RAM, i5 8 core processor and 256 GB disk capacity. 
In addition to the evaluation of the proposed heart sound denoising architecture, the performance of the proposed architecture has been compared with a wavelet-thresholding (WT) \cite{ghanbari2006new} based SoA technique and a baseline denoising auto-encoder (DAE) architecture \cite{banerjee2022noise},\cite{vincent2008extracting}.  respectively.
\begin{equation}\label{eq:RMSE}
RMSE = \sum_{n=1}^{K}\sqrt{(P_c[n]-P_p[n])^{2}}
\end{equation}

\begin{equation}\label{eq:ME}
    ME = median_{n=1}^{K}(|P_c[n]-P_p[n]|)
\end{equation}

\begin{equation}\label{eq:SNR}
SNR = 10log_{10}(\frac{\sum_{n=1}^{M}(P_c[n]-\mu_0)^{2}}{\sum_{n=1}^{M}(P_c[n]-P_p[n])^{2}})
\end{equation}
where $P_c$ is the clean PCG signal (target), $P_p$ is the noisy PCG signal and $\mu_{0}$ is the mean of $P_c$.

\begin{table}[]
	\caption{Comparative evaluation of the proposed denoising technique with existing state of the art denoising approaches}
	\label{performance}\centering
\begin{tabular}{p{3cm}p{1.25cm}p{1.25cm}p{1.25cm}}\hline
                               Method               & Mean RMSE & Mean MAE & Mean SNR      \\ \hline  
Proposed                    & 0.7588                            & 0.1063       & -2.4449  \\ \hline  
WT based            & 0.8772                            & 0.1232       & -3.6469  \\ \hline  
Baseline DAE based  & 1.8818                            & 0.2130       & -18.5644 \\ \hline  
\end{tabular}
\end{table}

\begin{figure}
	\begin{subfigure}{.48\textwidth}
	\centering
	\includegraphics[width=0.96\textwidth]{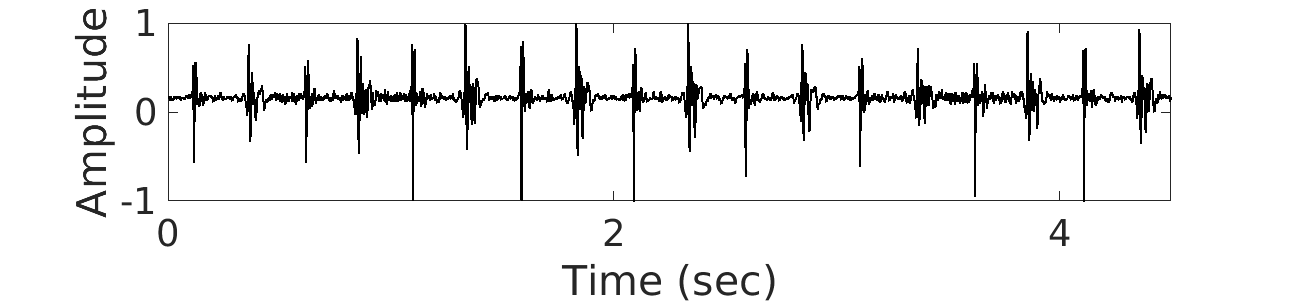}
	\caption{Reference clean heart sound signal}
	\label{fig:Result_Clean}
\end{subfigure}
	\begin{subfigure}{.48\textwidth}
	\centering
	\includegraphics[width=0.96\textwidth]{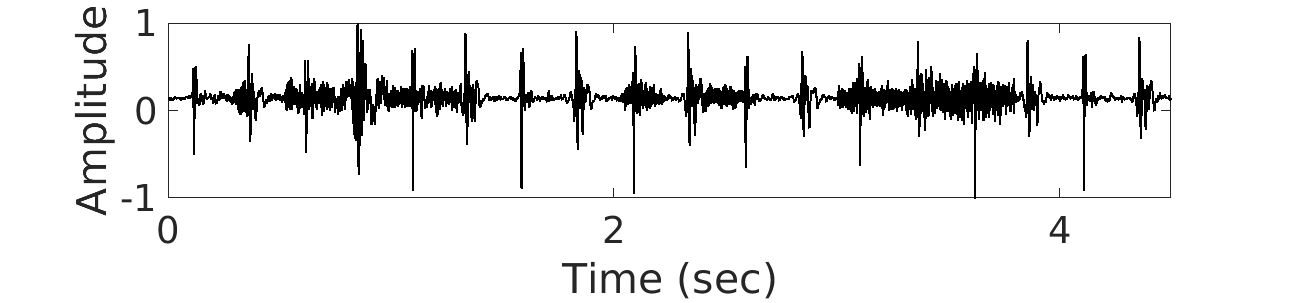}
	\caption{Noisy heart sound signal}
	\label{fig:Result_Noisy}
\end{subfigure}
	\begin{subfigure}{.48\textwidth}
		\centering
		\includegraphics[width=0.96\textwidth]{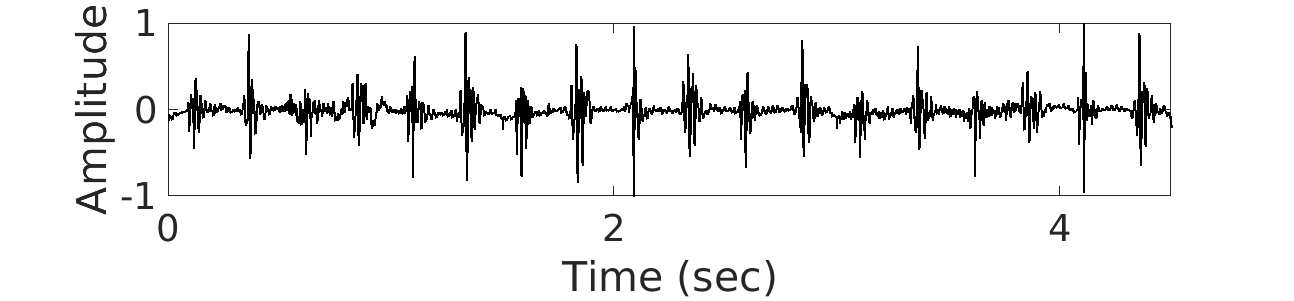}
		\caption{Heart sound signal denoised using the proposed denoising architecture}
		\label{fig:Result_Proposed}
	\end{subfigure}
	\begin{subfigure}{.48\textwidth}
	\centering
 	\includegraphics[width=0.96\textwidth]{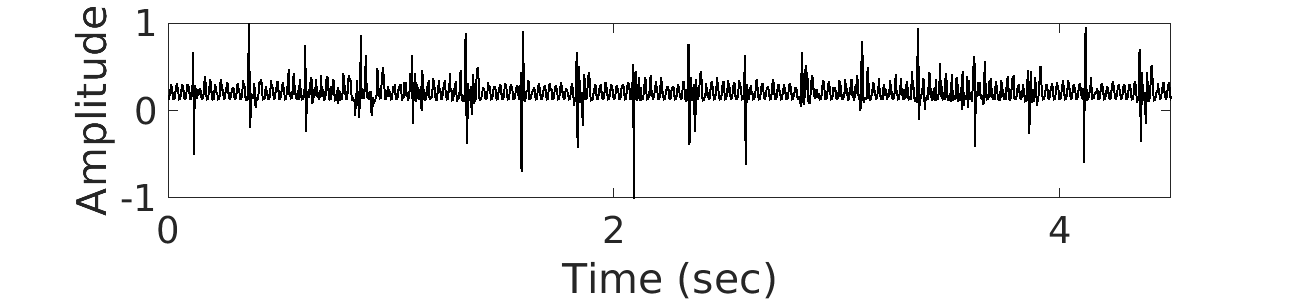}
	\caption{Heart sound signal denoised using baseline Wavelet thresholding based approach }
	\label{fig:Result_WT}
	\includegraphics[width=0.96\textwidth]{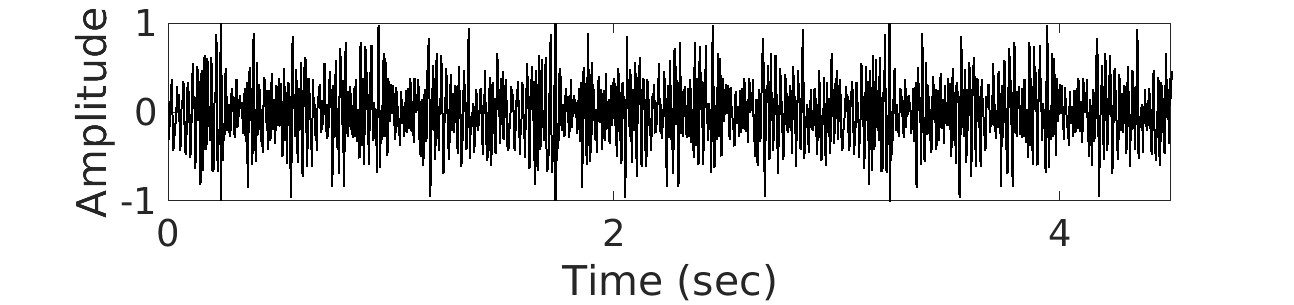}
	\caption{Heart sound signal denoised using baseline autoencoder-decoder architecture}
	\label{fig:Result_DAE}
\end{subfigure}
\caption{Performance plots of different noise cleaning architecture on a sample of heart sound recordings}
\label{fig:fig}
\end{figure}

The quantitative evaluation of the proposed denoising algorithm is done based on three metrics: root mean square error (RMSE), median absolute error (MAE) and signal to noise ratio (SNR). The mathematical representations of the three metrics are given in (\ref{eq:RMSE}), (\ref{eq:ME}) and (\ref{eq:SNR}). 
The performance of the proposed algorithm and the two SoA techniques in terms of the three metrics are reported in Table \ref{performance}. From the reported metric values it can be observed that the proposed denoising architecture has performed better than the two SoAs comprehensively. 
In addition, for qualitative assessment, a sample of the test noisy heart sound signal  and clean heart sound signal (target) are plotted in Fig. \ref{fig:Result_Noisy} and Fig.~\ref{fig:Result_Clean} respectively. The de-noised heart sound signals as obtained from the proposed denoising architecture is plotted in Fig.~\ref{fig:Result_Proposed}. Further, the denoised time series as obtained from the SoAs (WT and DAE) are plotted in Fig.~\ref{fig:Result_WT} and Fig.~\ref{fig:Result_DAE} respectively. It can be observed that the proposed methodology is able to effectively remove the real-world noises from the noisy signal while preserving the S1 and S2 characteristics. The wavelet based approach has introduced a narrow band of noise along the time series as well as thinned out the S1, S2 peaks thereby impacting its audio characteristics. The DAE based method has failed to perform any effective noise cleaning. 
\section{CONCLUSIONS}
Heart sounds signal are in general vulnerable to ambient noises and hence denoising is considered as a critical pre-processing step in the subsequent analysis for potential disease diagnosis. Simulating such noise-contaminated PCG signal is challenging and the existing denoising approaches use AWG as the noise source for data corruption. A machine learning model trained on such data is less likely to generalize in practical scenario. In this paper, we propose a novel pipeline for simulating realistic noisy PCG signals. Further, we propose a novel U-Net based PCG denoising algorithm that can reliably reconstruct both the amplitude and the phase information of the PCG data from realistic noisy PCG recordings. The realistic noisy PCG signals synthesized has been used to training the proposed deep learning model. Our experiments on publicly available dataset and subsequent quantitative and qualitative analysis and comparison with other existing SoA denoising algorithms clearly indicates the efficacy of the proposed approach. 
\newline Denoising is a critical application for any PCG recording device front-end. Therefore, our future works would be to optimize the model for effective deployment on low-powered embedded platforms. 
\bibliographystyle{unsrt}
\bibliography{ref_embc}
\end{document}